\documentclass[a4paper,fleqn]{cas-dc}

\usepackage[authoryear,longnamesfirst]{natbib}
\usepackage{lineno,hyperref}
\usepackage{ulem}
\usepackage{algorithm2e,setspace}
\usepackage{subfigure}
\usepackage{multirow}
\usepackage{bbding}
\usepackage{mathrsfs}
\usepackage{amsthm,amsmath,amssymb}
\usepackage{tabularx}
\usepackage{natbib}

\def\tsc#1{\csdef{#1}{\textsc{\lowercase{#1}}\xspace}}
\tsc{WGM}
\tsc{QE}
\tsc{EP}
\tsc{PMS}
\tsc{BEC}
\tsc{DE}

\begin{document}
\let\WriteBookmarks\relax
\def\floatpagepagefraction{1}
\def\textpagefraction{.001}

\shorttitle{Unraveling Complex Data Diversity in Underwater Acoustic Target Recognition through CMoE}

\shortauthors{Xie et~al.}

\title [mode = title]{Unraveling Complex Data Diversity in Underwater Acoustic Target Recognition through Convolution-based Mixture of Experts}                      




%
\author[address1,address2]{Yuan Xie}[
                        style=chinese,
                        orcid=0000-0003-3803-0929]


\ead{xieyuan@hccl.ioa.cn.cn}



\credit{Conceptualization, Methodology, Software, Validation, Formal analysis, Investigation, Data Curation, Writing - Original Draft, Writing - Review \& Editing, Visualization}

\author[address1,address2]{Jiawei Ren}[style=chinese]
\ead{renjiawei@hccl.ioa.cn.cn}
\credit{Methodology, Investigation, Data Curation, Supervision}

\author[address1,address2,address3]{Ji Xu}[style=chinese, orcid=0000-0002-3754-228X]
\ead{xuji@hccl.ioa.cn.cn}
\cormark[1]

\credit{Resources, Writing - Review \& Editing, Supervision, Project administration, Funding acquisition}

\affiliation[address1]{organization={Key Laboratory of Speech Acoustics and Content Understanding, Institute of Acoustics, Chinese Academy of Sciences},
    addressline={No.21, Beisihuan West Road, Haidian District},
    postcode={100190},
    city={Beijing},
    country={China}}
    
\affiliation[address2]{organization={University of Chinese Academy of Sciences},
    addressline={No.80, Zhongguancun East Road, Haidian District}, 
    postcode={100190},
    city={Beijing},
    country={China}}

\affiliation[address3]{organization={State Key Laboratory of Acoustics, Institute of Acoustics, Chinese Academy of Sciences},
    addressline={No.21, Beisihuan West Road, Haidian District},
    postcode={100190},
    city={Beijing},
    country={China}}

\cortext[cor1]{Corresponding author}

\fntext[fn1]{The source code of this paper could be obtained from https://github.com/xy980523/UATR-CMoE.}


\begin{abstract}
Underwater acoustic target recognition is a difficult task owing to the intricate nature of underwater acoustic signals. The complex underwater environments, unpredictable transmission channels, and dynamic motion states greatly impact the real-world underwater acoustic signals, and may even obscure the intrinsic characteristics related to targets. Consequently, the data distribution of underwater acoustic signals exhibits high intra-class diversity, thereby compromising the accuracy and robustness of recognition systems. 
To address these issues, this work proposes a convolution-based mixture of experts (CMoE) that recognizes underwater targets in a fine-grained manner. The proposed technique introduces multiple expert layers as independent learners, along with a routing layer that determines the assignment of experts according to the characteristics of inputs. This design allows the model to utilize independent parameter spaces, facilitating the learning of complex underwater signals with high intra-class diversity. Furthermore, this work optimizes the CMoE structure by balancing regularization and an optional residual module. To validate the efficacy of our proposed techniques, we conducted detailed experiments and visualization analyses on three underwater acoustic databases across several acoustic features. The experimental results demonstrate that our CMoE consistently achieves significant performance improvements, delivering superior recognition accuracy when compared to existing advanced methods.
\end{abstract}



\begin{keywords}
Underwater acoustic target recognition \sep Deep learning \sep Convolutional neural network \sep Adaptive route assignment \sep Mixture of experts
\end{keywords}

\maketitle

\section{Introduction}
\subsection{Background}
Underwater acoustic target recognition plays a crucial role in the field of marine acoustics. The goal of this task is to automatically analyze the sound radiated by underwater targets and predict the type of underwater targets. The long detection range, reassuring concealment, and low deployment cost make it indispensable in practical applications. This technology finds extensive use in underwater surveillance, marine resources development and protection, as well as security defense~\citep{irfan2021deepship,jia2022deep,sutin2010stevens}.

In response to the growing need for robust underwater acoustic recognition systems, numerous research efforts have been dedicated to this area in recent years~\citep{li2017denoising,ke2020integrated,simonovic2021acoustic,xie2022underwater}. The underwater acoustic recognition systems typically consist of two main components: acoustic feature extraction and recognition models. To extract discriminative acoustic features from raw underwater acoustic signals, researchers have proposed various acoustic analysis strategies, including Fourier transform~\citep{xie2022underwater,liu2021underwater}, Hilbert–Huang transform~\citep{wang2014robust}, wavelet transform~\citep{jia2022deep,xie2022adaptive}, Mel filtering~\citep{liu2021underwater,zhang2016feature}, LOFAR (low-frequency analysis recording)~\citep{chen2021underwater}, cepstrum extraction~\citep{zhang2016feature,das2013marine}, and more. These techniques leverage the acoustic properties of targets to extract low-dimensional acoustic features. Once the features are extracted, recognition systems employ various models, such as classic machine learning models~\citep{wang2014robust,das2013marine,erkmen2008improving} or deep neural networks~\citep{xie2022underwater,liu2021underwater,xie2022adaptive,zhang2021integrated,ren2022ualf,khishe2019passive}, to exploit the discriminative patterns in acoustic features and make predictions about the target categories.

Besides, underwater acoustic target recognition systems also heavily rely on underwater acoustic signal databases. The performance of recognition systems is greatly influenced by factors such as the scale, authenticity, and diversity of the utilized data.~\citep{irfan2021deepship}. Given the high cost and equipment dependency associated with collecting real-world underwater acoustic data~\citep{hovem2012marine}, previous research has predominantly utilized synthetic data~\citep{das2013marine}, simulated data~\citep{zhang2016feature}, or privately recorded data~\citep{wang2014robust}. With the increasing practical demand in recent years, open-source underwater databases have begun to be released, such as Shipsear~\citep{santos2016shipsear} and DeepShip~\citep{irfan2021deepship}. Large-scale databases offer researchers access to abundant and diverse real-world signals, which can contribute to enhancing the generalization capability of recognition systems in real-world scenarios.

The rapid advancements in deep learning algorithms~\citep{lecun2015deep} and the availability of open-source underwater acoustic databases have propelled deep neural networks to the forefront of recognition systems in recent years~\citep{xie2022underwater,saffari2023fuzzy,khishe2022drw}. Deep neural networks leverage their complex topological structures and numerous nonlinear operators to support highly intricate modeling, while open-source databases provide an ample amount of training data for these networks. In comparison to classical methods, which exhibit limitations when dealing with large-scale data characterized by diverse feature spaces~\citep{irfan2021deepship}, recognition systems based on deep learning showcase a significant performance advantage~\citep{xie2022adaptive,xie2022underwater} on existing databases.

\subsection{Motivation}

\begin{figure*}
    \centering
    \includegraphics[width=0.7\linewidth]{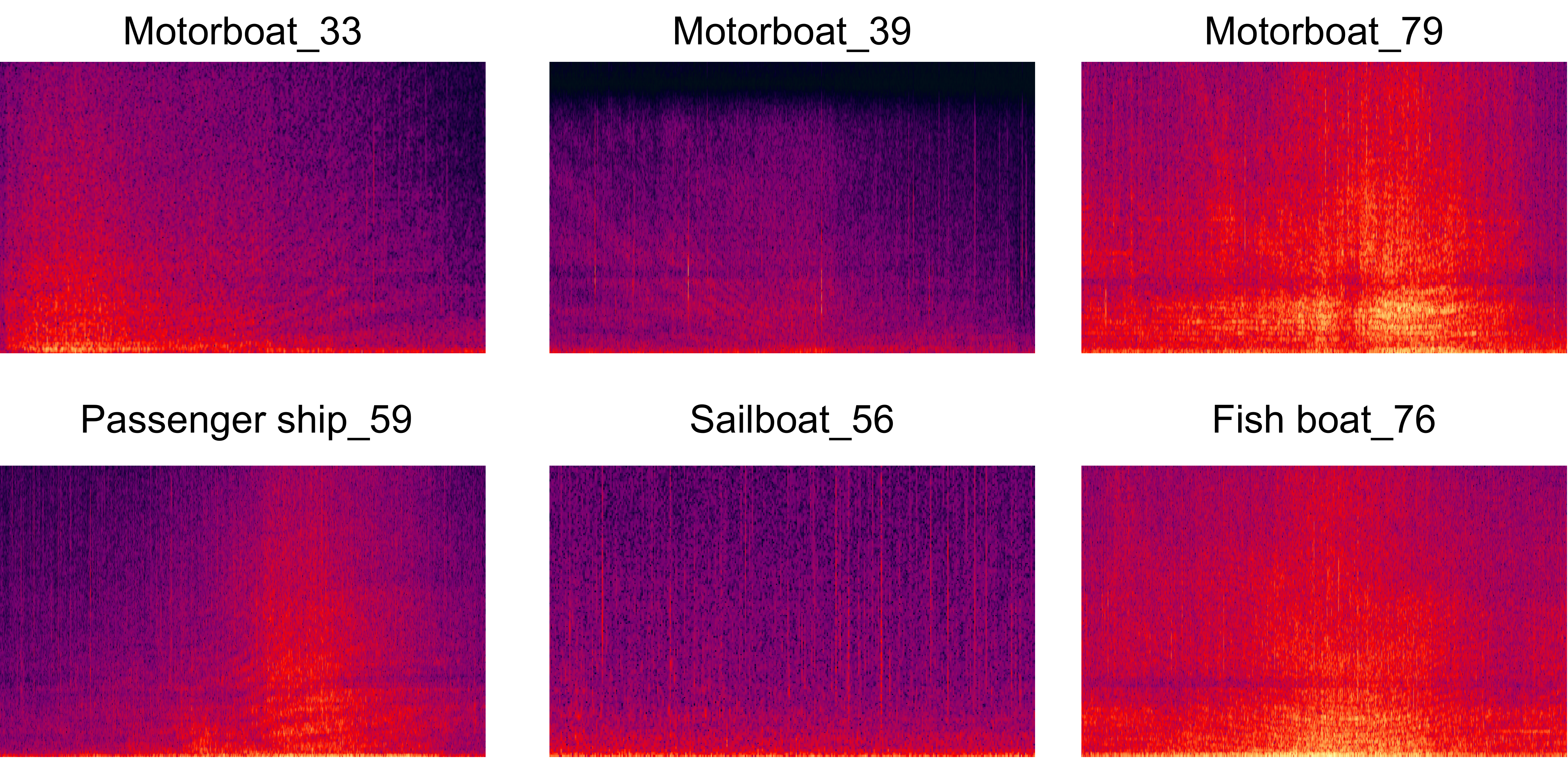}
    \caption{Spectrograms of several samples in the Shipsear dataset. Motorboat\_33 records the start and stop of the motorboat ``Dud''; Motorboat\_39 records the arrival of the motorboat ``Dud''; Motorboat\_79 records the passing of the motorboat ``Zodiac''; Passenger ship\_59 records ``Marde Mouro'' sailing towards the port with considerable speed; Sailboat\_56 records a sailboat passing in a very close distance; Fishboat\_76 records the fish boat passing.}
    \label{fig1}
    \vspace{-5px}
\end{figure*}

Despite notable advancements on open-source datasets, the performance of recognition models in practical underwater scenarios remains unsatisfactory ~\citep{xie2022underwater}. This can be attributed to the complexity of underwater acoustic signals, which are influenced by various interference factors, including intricate underwater environments, unpredictable transmission channels, and volatile vessel motion states~\citep{erbe2019effects}. Besides, variations in propeller and engine technology further impact the acoustic characteristics of radiated noise~\citep{khishe2020classification}. These interference factors contribute to the complexity and indistinguishability of collected signals, leading to a high intra-class diversity within the overall data distribution. Figure~\ref{fig1} visually illustrates the spectrogram comparisons of several samples from the Shipsear dataset, clearly demonstrating significant intra-class diversity in spectrograms among motorboats with different motion states (e.g., No.33 indicating start and stop and No.39 indicating arrival) and types (e.g., No.33 representing the motorboat ``Dud'' and No.79 representing the motorboat ``Zodiac''). According to related studies~\citep{schutz2013k}, intra-class diversity can result in potential misrecognition, particularly when data is limited.

Moreover, underwater targets (e.g., vessels) often share similar vibrational modes, such as propeller cavitation noise, and rhythm modulation noise caused by diesel piston movement. This similarity can lead to commonalities among different target categories. As observed in the spectrograms presented in Figure~\ref{fig1}, passenger ships, sailboats, and fish boats exhibit a certain degree of similarity with motorboats. Addressing this inter-class similarity requires models to delve into high-level semantic concepts with discernible characteristics, which introduces the risk of overfitting and further complicates the recognition task.

Currently, existing literature rarely focuses on the distribution characteristics of intra-class diversity and inter-class similarity in underwater acoustic signals. For such complex signals, a natural idea is to increase the number of parameters of the model to support complex modeling. However, the scarcity of underwater data imposes significant limitations as complex models can exacerbate the overfitting phenomenon.

\subsection{Our Work}
In this work, we present innovative techniques to address the aforementioned challenges in underwater acoustic recognition. To mitigate the impact of intra-class diversity, it is necessary to develop an adaptive model capable of effectively processing diverse data. Drawing inspiration from successful applications of the mixture of experts (MoE) paradigm in computer vision~\citep{ahmed2016network,riquelme2021scaling} and natural language processing~\citep{fedus2021switch,xie2022moec}, we propose the convolution-based mixture of experts (referred to as CMoE) for underwater acoustic target recognition. CMoE incorporates multiple expert layers and a routing layer, where the routing layer dispatches inputs to the most suitable expert layer based on high-level representations. This approach enables adaptive disassembly of diverse data, allowing the model to learn underwater acoustic signals with multiple independent parameter spaces. Furthermore, to address the issue of inter-class similarity, we position the expert layers in the final layer of the model. It allows expert layers to learn from high-level semantic concepts and focus on discriminative characteristics within data that exhibit inter-class similarity.

Additionally, we introduce balancing regularization to handle the load balance problem and optimize our CMoE structure by incorporating an optional residual module. Experiments demonstrate that our CMoE can consistently achieve superior performance across various underwater acoustic databases. The contributions of our work can be summarized as follows:

\begin{itemize}
\item we reveal the unique characteristics of underwater acoustic signals, including intra-class diversity and inter-class similarity, along with the limitations of existing approaches;

\item we propose the convolution-based mixture of experts to unravel complex data diversity, which captures latent characteristics from high-level representations and adaptively learns diverse data with multiple independent parameter spaces;

\item we optimize our CMoE structure through balancing regularization and the incorporation of an optional residual module;

\item we provide detailed visualizations and corresponding analyses of the recognition results and the routing assignment of expert layers.
\end{itemize}



\section{Related Works}
\subsection{Underwater Acoustic Target Recognition}
The research on underwater acoustic target recognition primarily revolves around acoustic feature extraction and applications of recognition algorithms. Early studies employed classic machine learning techniques to process manually designed low-dimensional acoustic features. For instance, Das et al.~\citep{das2013marine} utilized cepstral features with cepstral liftering and Gaussian mixture models (GMMs) for marine vessel classification; Wang and Zeng~\citep{wang2014robust} employed bark-wavelet analysis in combination with the Hilbert-Huang transform to analyze signals, and employed support vector machines (SVMs) as the classifier. Moreover, cepstrum-based acoustic features from the audio and speech domains, such as Mel frequency cepstrum coefficients (MFCCs), have also yielded promising results in ship-radiated noise recognition tasks~\citep{zhang2016feature,khishe2019passive}. However, recent studies have revealed limitations in the general applicability of recognition systems in complex underwater scenarios when relying solely on manually designed low-dimensional acoustic features~\citep{irfan2021deepship,xie2022underwater}. Moreover, classical machine learning models may struggle to achieve satisfactory performance when confronted with large-scale data with diverse feature spaces~\citep{irfan2021deepship}. As a result, recognition systems based on these classic paradigms face challenges in accurately recognizing unseen data in practical ocean scenarios.

With the development of deep learning~\citep{lecun2015deep} and the accumulation of open-source underwater acoustic databases~\citep{irfan2021deepship,santos2016shipsear}, recognition algorithms based on deep neural networks have gained prominence. As reported in the literature, Zhang et al.~\citep{zhang2021integrated} utilized the short-time Fourier transform (STFT) amplitude spectrum, STFT phase spectrum, and bispectrum features as inputs for convolutional neural networks; Liu et al.~\citep{liu2021underwater} employed convolutional recurrent neural networks with 3-D Mel-spectrograms and data augmentation for underwater target recognition; Xie et al.~\citep{xie2022adaptive} utilized learnable fine-grained wavelet spectrograms with the deep residual network (ResNet)~\citep{he2016deep} to adaptively recognize ship-radiated noise; Ren et al.~\citep{ren2022ualf} employed learnable Gabor filters and ResNet for constructing an intelligent underwater acoustic classification system. In contrast to classical machine learning paradigms, deep learning methods often favor acoustic features that encapsulate comprehensive information, such as time-frequency spectrograms~\citep{liu2021underwater,xie2023guiding,xu2023underwater}. The large number of parameters and complex nonlinearities in neural networks enable the model to effectively exploit information contained within comprehensive features. The superior performance on available datasets has propelled these methods to become the mainstream approach for underwater acoustic target recognition. However, deep learning-based systems exhibit limited robustness in practical application scenarios characterized by scarce and complex data. In this regard, many studies have employed techniques like denoising~\citep{li2017denoising,ghavidel2022sonar} and data augmentation\citep{liu2021underwater,chen2022underwater,xu2023underwater} to enhance the robustness of recognition systems by improving data quality or quantity. According to our research, we are the first to address the intra-class diversity and inter-class similarity of underwater signals, and take it as the starting point for building a robust recognition system.

\subsection{Mixture of Experts}
To efficiently and effectively recognize complexly distributed underwater signals, this work draws inspiration from the Mixture of Experts (MoE) approach. MoE enables discriminative processing of diverse-distributed data while minimizing the introduction of excessive parameters that could lead to overfitting. The original formulation of MoE models was introduced by Jacobs et al.~\citep{jacobs1991adaptive}, including a variable number of expert models and a single gate to combine their outputs. Subsequent work by Collobert et al.~\citep{collobert2002parallel,collobert2003scaling} applied the MoE concept to classic machine learning algorithms like support vector machines. In recent years, Shazeer et al.~\citep{shazeer2017outrageously} explored the transition to conditional computing through sparse expert activations, wherein a fixed number of experts were activated in LSTMs. Fedus et al.~\citep{fedus2021switch} extended the MoE structure to large-scale Transformers~\citep{vaswani2017attention}, revealing its potential to construct scalable models with reasonable overhead. As research on MoE deepened, various studies applied it to scale up Transformers~\citep{fedus2021switch,xie2022moec,rajbhandari2022deepspeed,riquelme2021scaling} and convolutional neural networks~\citep{gross2017hard,ahmed2016network,wang2020deep}. The sparse-gated MoE significantly increases model capacity while incurring minimal compute overhead, leading to remarkable achievements in natural language processing and computer vision. In this work, we propose a specialized convolutional MoE with sparsely-activated experts for underwater acoustic target recognition, making us the first, to the best of our knowledge, to apply the mixture of experts in this domain.

\section{Methodology}
This section begins by presenting the acoustic feature extraction methods employed in this work. Following that, we introduce the front-end backbone network, the expert layers, the routing layer, and the optional residual module of CMoE, respectively. Finally, we introduce the balanced regularization strategy adopted to address the load imbalance problem inherent in the MoE structure.

\subsection{Acoustic Feature Extraction}
In order to validate the generalizability of our proposed strategies, we employ four feature extraction techniques in this study. For raw signals, we begin by computing the spectrums through framing, windowing, and short-time Fourier transform (STFT). Subsequently, the real component is extracted and integrated across the time dimension to derive the STFT spectrogram. Following this, we employ Mel (Bark) filter banks to perform the filtering on the framed spectrums. Equation (1) illustrates that the Mel (Bark) filter bank comprises a set of bandpass filters, distributed based on the non-linear Mel (Bark) scale. Notably, the Mel (Bark) filter banks exhibit higher density at low frequencies to achieve enhanced frequency resolution. Finally, the filtered spectrums undergo conversion to Mel (Bark) spectrograms through a logarithmic scale and the integration across the time dimension.

\begin{equation}
\begin{aligned}
    &Mel(f)
    =2595\times log(1+\frac{f}{700}),\\
    &Bark(f)
    =6 \times arsinh(\frac{f}{600}).
\end{aligned}
\end{equation}

Furthermore, we acquire the CQT spectrogram by conducting the \textbf{C}onstant \textbf{Q} \textbf{T}ransform. This involves convolving the spectrum of each frame, obtained from STFT, with the CQT kernel. The CQT kernel consists of a bank of bandpass filters that are logarithmically spaced in frequency. The $k-$th frequency component $f_k$ can be formalized as depicted in Equation (2), where $b$ represents the octave resolution, and $f_{max}$ and $f_{min}$ denote the maximum and minimum frequencies to be processed, respectively.

\begin{equation}
    CQT(f_k)=2^{k/b} f_{min}. \quad f_{min}\leq f_k\leq f_{max}.
\end{equation}

Then, we integrate the magnitude of the filtered spectrum across the time dimension to yield the CQT spectrogram. Notably, the CQT spectrogram exhibits higher temporal resolution at high frequencies.

\subsection{Front-end Backbone Network}

\begin{figure*}
    \centering
    \includegraphics[width=0.8\linewidth]{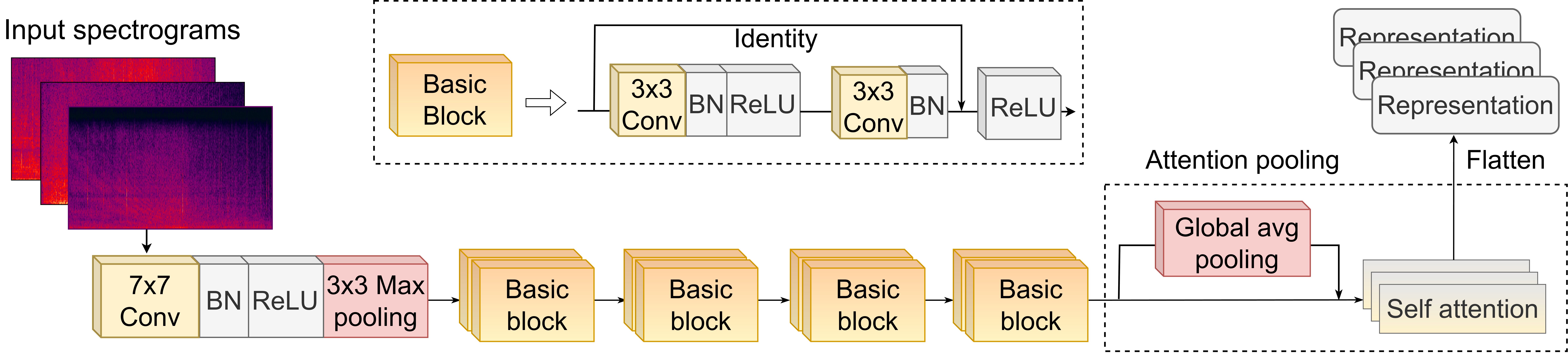}
    \caption{The model structure of the front-end backbone model - ResNet with attention pooling. ``Conv'' represents the convolutional layer, and ``BN'' represents the batch normalization layer.}
    \label{fig2}
    \vspace{-2px}
\end{figure*}

\begin{table}[ht]
\normalsize
    \centering
    \caption{Specific network structure for our CMoE, including the front-end backbone, the expert layer and the routing layer. ``num class'' represents the number of classes to be predicted and ``num experts'' represents the number of expert layers.}
	\scalebox{0.65}{\begin{tabular}{ll}
		\hline
		  Module&Specific network layer\\ 
            \hline
            Front-end backbone&Conv2d(1, 64, kernel size=7, stride=2, padding=3) \\
            &Batch Normalization 2d(num features=64)\\ 
             &ReLU()\\ 
             &Max pooling(kernel size=3, stride=2, padding=1)\\
             &Basic block(64,64),Basic block(64,64)\\
             &Basic block(64,128),Basic block(128,128)\\
             &Basic block(128,256), Basic block(256,256)\\
             &Basic block(256,512), Basic block(512,512)\\
             &Attention pooling(output size=(1, 1))\\
             
            \hline
            Basic block(in dim, out dim)& Conv2d(in dim, out dim, kernel size=3, padding=1)\\
            & Batch Normalization 2d(num features=out dim)\\
            &ReLU()\\ 
            & Conv2d(out dim, out dim, kernel size=3, padding=1)\\
            & Batch Normalization 2d(num features=out dim)\\
		\hline
            Expert layer&Linear(in features=512, out features=128)\\
            &Batch Normalization 1d(num features=128)\\
            &ReLU()\\
            &Linear(in features=128, out features=num class)\\
            \hline
            Routing layer&Linear(in features=512, out features=num experts)\\
            \hline
        \label{tab_structure}
	\end{tabular}}
\end{table}

Following the feature extraction stage, we present the overall process and structure of our proposed CMoE model. Our CMoE comprises two main components. The first component is the front-end backbone network, responsible for transforming the input acoustic features into fixed-dimensional representations.

In this study, we adopt the optimized deep residual network, ResNet with attention pooling~\citep{he2016deep,wang2018non}, as our front-end backbone. This choice is based on its superior recognition performance, as demonstrated in our preliminary experiments (see Section 5.1). The detailed structure of the front-end backbone network is illustrated in Table~\ref{tab_structure} and Figure~\ref{fig2}. It consists of a convolution layer stacked with a batch normalization (BN) layer, a ReLU layer, a max-pooling layer, followed by four residual layers and an attention pooling layer. Each residual layer comprises two basic blocks stacked together. The structure of the basic block, enclosed in a dotted box in Figure~\ref{fig2}, includes convolution layers, BN layers, ReLU layers, and a skip connection. Mathematically, the basic block can be defined as:

\begin{equation}
    y=F(x, \{W_i\}) + x,
\end{equation}

where $x$ and $y$ represent the input and output vectors of the basic blocks, respectively. The function $F = W_2 \sigma(W_1x)$, with $\sigma$ denoting the ReLU layer and $W_1, W_2$ representing the learnable mappings for the two convolutional layers. Following the residual layers, we incorporate an attention pooling layer and a flattening operation to obtain the output representations $\sim \mathbb{R}^{batch size \times 512}$. The attention pooling layer incorporates global average pooling and multi-head self-attention operations, allowing for the assignment of dynamic weights to different regions of the feature map. This approach helps in focusing on useful information and improving the quality of the representations. The output representations from the front-end backbone network are then passed to subsequent layers in the network.

\subsection{Expert Layer, Routing Layer, and Residual Module}

\begin{figure*}
    \centering
    \includegraphics[width=0.7\linewidth]{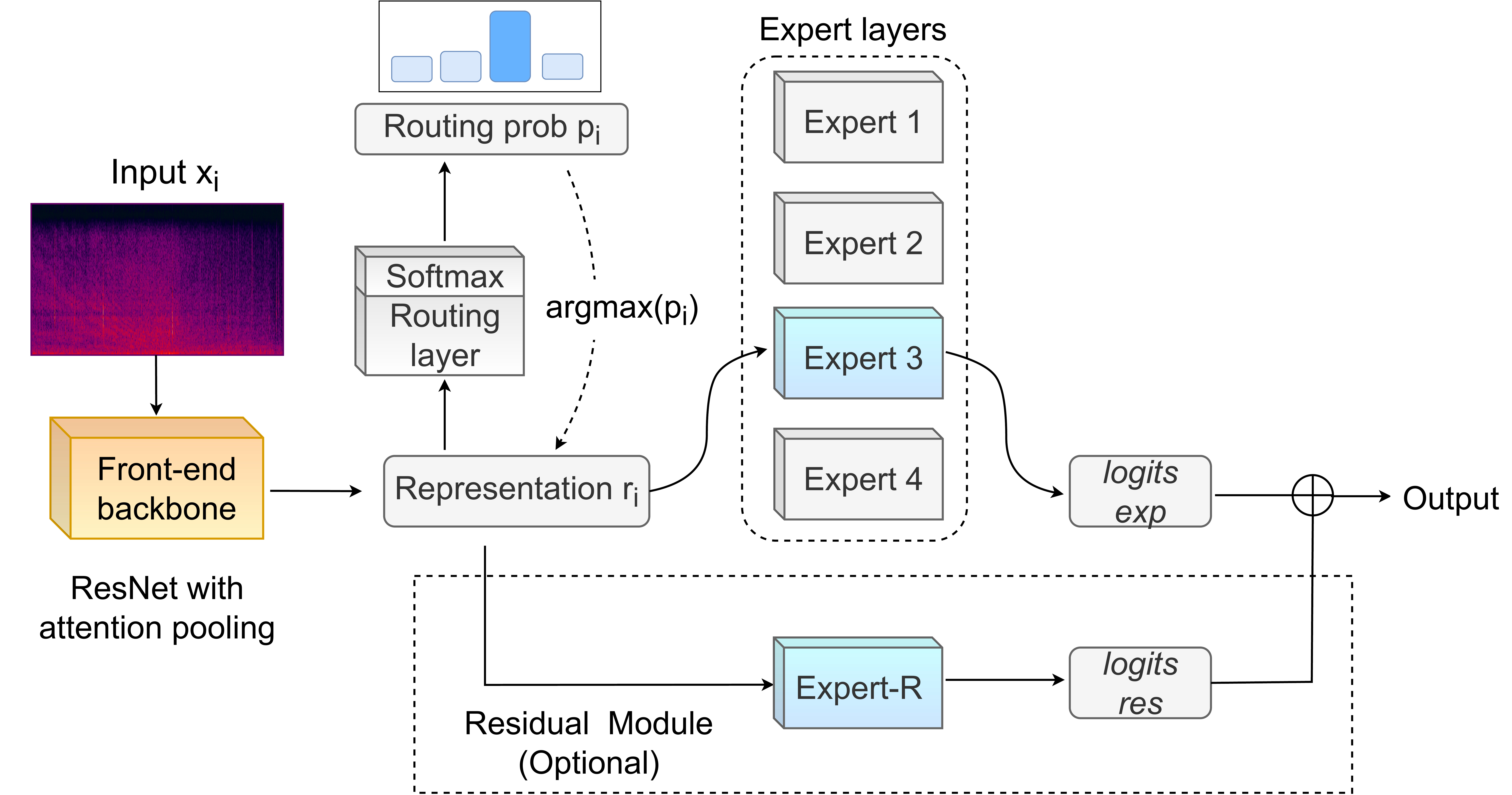}
    \caption{The overall process of our proposed CMoE, including routing probability calculation, expert assignment, and optional residual module.}
    \label{fig3}
    \vspace{-2px}
\end{figure*}

The structure and training pipeline of the expert layers and the routing layer are presented in Table~\ref{tab_structure} and Figure~\ref{fig3} respectively. Both the expert layers and the routing layer take the output representations of the front-end backbone network as input. The expert layer consists of a multi-layer perceptron with two linear layers (see Table~\ref{tab_structure}). Each expert has a consistent structure, but its parameters are independent. Activation of each expert occurs only when suitable input is encountered, facilitating fine-grained and differentiated learning of diverse underwater data. The routing layer consists of a simple linear layer that adaptively guides the assignment of inputs by calculating the routing probability. This design enables the model to disassemble diverse underwater acoustic data using multiple independent parameter spaces, thereby reducing the impact of intra-class diversity. The following paragraph provides a formulaic description of the detailed process.

Let's denote the batch size as $n$, input spectrograms as $\{x_i\}$, the corresponding label as $\{y_i\}$ (i=1,2...$n$), the number of expert layers as $m$, and the expert layers as $E_1(\cdot),...E_m(\cdot)$. The front-end backbone model first takes $x_i$ as input and produces high-level representations $r_i\sim \mathbb{R}^{n \times 512}$. These representations $r_i$ are then fed into the linear routing layer $G(\cdot)$, sequentially obtaining the routing score $s_i = G(r_i)\sim \mathbb{R}^{1 \times m}$. Then, we apply the softmax function\footnote{As for the choice of normalized function, we have carried out relevant comparative experiments in Section 5.5. The introduction here uses the softmax function by default.} to normalize the routing score $s_i$ into the routing probability $p_i\sim \mathbb{R}^{1 \times m}$, which indicates the probability of dispatching the sample $x_i$ to each expert.

\begin{equation}
\begin{aligned}
    &p_i = softmax(G(r_i)), \quad i\epsilon [1,n] \\
    &logits_{exp} = E_{argmax(p_i)}(r_i).
\end{aligned}
\end{equation}

Then, the model sends the representation $r_i$ to the expert layer with the highest corresponding probability value. The overall routing assignment process is described in Equation (4). Take the CMoE composed of 4 experts as an example, when $p_i$=(0.1, 0.2, 0.4, 0.3), the corresponding representation $r_i$ should be sent to $E_3(\cdot)$ with the maximum probability.

Furthermore, to prevent overfitting from affecting routing assignment and model performance, we adopt the concept of residual connection and propose \textbf{R}esidual CMoE (RCMoE). As depicted in Figure~\ref{fig3} (the dashed box at the bottom), RCMoE retains the structure of CMoE while adding an additional fixed expert layer $E_R(\cdot)$. The representation $r_i$ is directly fed into the fixed expert layer without performing routing calculations, generating $logits_{res}=E_R(r_i)$. Finally, $logits_{res}$ is added to the output of the expert layers $logits_{exp}$ to obtain the overall logits. The optimization goal is to minimize the cross-entropy loss between the overall logits and labels $\{y_i\}$. We provide additional pseudocode in Algorithm 1 to help better understand the training flow of CMoE and RCMoE.

\setlength{\algomargin}{1em}
\begin{algorithm}[t]
\caption{Training flow of our (R)CMoE. }
\label{alg:learning}
\LinesNumbered
\SetKwFor{For}{for}{do}{end}
\KwData{input spectrograms $\mathcal{X}=\{x_i\}$; label $\mathcal{Y}=\{y_i\}$.}
\KwIn{number of experts $m$; batch size $n$; front-end model $F$; routing layer $G$; experts $E_1(\cdot),...E_m(\cdot)$; fixed expert $E_R(\cdot)$.}
\While{not done}{
    Sample batches $(x_i, y_i) \sim (\mathcal{X, Y})$. \ $i=1,2,...,n$\\
    \For{$(x_i, y_i)$}{
        \# compute representations \\
        $r_i = F(x_i)$\\
        \# compute routing probabilities \\
        $p_i= softmax(G(r_i), dim=-1)$\\
        \# dispatch $r_i$ to experts \\
        \For{$1\leq i \leq n$}{
        $logits\ exp[i,:] = E_{argmax(p_i)}(r[i,:])$\\
        $logits\ res[i,:] = E_R(r[i,:])$
        }
        \# compute final logits \\
        $logits$ = $logits\ exp$ for CMoE \\
        
        $logits$ = $(logits\ exp+logits\ res)$ for RCMoE
    }
    Update weights with loss $\mathcal{L}_{CE}$ = Cross Entropy$(logits,\{y_i\}).$
}
\end{algorithm}

\subsection{Balancing Regularization}
During training, a significant issue of load imbalance arises among experts, as noted by Shazeer et al.~\citep{shazeer2017outrageously}. This imbalance manifests when a small number of experts receive the majority of inputs, while many other experts remain inadequately trained. To address this concern, we adopt the balancing regularization approach proposed by Fedus et al.~\citep{fedus2021switch}, aiming to promote a more equitable distribution of the workload across all experts. We inherit the notation used in the previous subsection, the balance loss is computed as follows:

\begin{equation}
    \mathscr{L}_{balance}=\alpha m \cdot \sum_{j=1}^{m} ef_{j} \cdot ep_{j},
\end{equation}

where $ef_j$ is represents the fraction of inputs dispatched to the $j$-th expert. We denote the number of inputs dispatched to the $j$-th expert as $Count_j$, thus $ef_j$ can be calculated as $ef_j = \frac{Count_j}{n}$. Additionally, $ep_j$ corresponds to the average routing probability dispatched to the $j$-th expert within the batch. It is determined by averaging the routing probabilities across the batch:

\begin{equation}
    ep_{j} = \frac{1}{n} \sum_{i=1}^{n} p_i[j],
\end{equation}

where $p_i[j]$ denotes the routing probability of dispatching token $x_i$ to the $j$-th expert. The balancing regularization term in Equation (5) encourages uniform routing, as minimizing it favors a uniform distribution. To control the impact of the balancing regularization during training, we introduce a hyper-parameter $\alpha$ as a coefficient for the regularization term. In this work, we set $\alpha = 10^{-2}$ as the default value, striking a balance between ensuring load balancing and avoiding excessive interference with the primary cross-entropy objective. Thus, the overall training objective can be presented as follows:

\begin{equation}
\begin{aligned}
    \mathcal{L}
    &=\mathcal{L}_{CE} + \mathcal{L}_{balance}\\
    &=Cross\ Entropy(logits,y) + \alpha m \cdot \sum_{j=1}^{m} ef_{j} \cdot ep_{j}.
\end{aligned}
\end{equation}

\section{Experiment Setup}
\subsection{Datasets}
In this work, we utilize three distinct datasets of underwater ship-radiated noise at varying scales. The detailed information is provided in Table~\ref{tab1} and the subsequent paragraphs.

1. Shipsear~\citep{santos2016shipsear} is an open-source database of underwater recordings of ship and boat sounds. The database comprises 90 records from 11 different vessel types, totaling nearly three hours of duration. To ensure an adequate amount of records for the ``train, validation, test'' split, we have selected a subset of 9 categories (dredger, fish boat, motorboat, mussel boat, natural noise, ocean liner, passenger ship, ro-ro ship, sailboat) from the Shipsear database for the recognition task.

2. Our private dataset~\citep{ren2019feature} - DTIL is collected from Thousand Island Lake, which contains multiple sources of interference. It contains 330 minutes of speedboat recordings and 285 minutes of experimental vessel recordings.

3. DeepShip~\citep{irfan2021deepship} is an open-source underwater acoustic benchmark dataset, which consists of 47.07 hours of real-world underwater recordings of 265 different ships belonging to four classes: cargo ship, passenger ship, tanker, and tugboat.

\subsection{Effective Frequency Bands}
To reduce the redundancy of acoustic features, we apply effective frequency bands as substitutes for full bands during feature extraction. It involves performing a time-frequency transformation within the bandwidth that encompasses the most relevant frequency components. By doing so, we aim to reduce redundant components in input features while simultaneously reducing time and hardware consumption. Considering that the useful energy of signals predominantly concentrates on distinct frequency ranges in Shipsear, DTIL, and DeepShip, we establish independent effective frequency bands for each dataset (refer to Table~\ref{tab1}). It is worth mentioning that the upper limit of the effective frequency band must be set below half of the sample rate according to the Nyquist theory. Further details about the experiments conducted to determine the optimal effective frequency band selection can be found in Section 5.1. Additionally, Table~\ref{tab1} also presents the dimensions of each acoustic feature on the three datasets.

\begin{table*}[ht]
\normalsize
    \centering
    \caption{Information of the three datasets. ``sr'' represents the sampling rate and ``dim'' represents the feature dimensions.}
	\scalebox{0.7}{\begin{tabular}{llcllccc}
		\hline
		  dataset  & duration (hours) & sr (Hz) & efficient band (Hz)&
            STFT dim&Mel dim&Bark dim&CQT dim\\
            \hline
            Shipsear  & 2.94 & 52734 & 100-26367&
            1200,1318&1200,300&1200,300&900,340\\
            DTIL  & 10.25 & 17067 & 100-2000&
            1200,99&1200,300&1200,300&900,230\\
            Deepship & 47.07 & 32000 & 100-8000&
            1200,400& 1200,300& 1200,300&900,290\\
		\hline
        \label{tab1}
	\end{tabular}}
\end{table*}

\subsection{Data Division}
In this work, each signal is cut into 30-second segments with a 15-second overlap. To prevent information leakage, we ensure that segments in the training set and the test set do not originate from the same audio track. This precautionary measure guarantees that the reported accuracy truly reflects the system's recognition ability and generalization performance, rather than its memory capacity.

We find that almost all previous works on underwater acoustic target recognition have not disclosed their train-test splits, making it challenging to establish fair comparisons. To address this issue, we provide our carefully selected train-test splits for Shipsear (see Table~\ref{tabD1}\footnote{The train-test split for Shipsear is also released at https://github.com/xy980523/ShipsEar-An-Unofficial-Train-Test-Split}) and DeepShip (see Table~\ref{tabD2}). 15\% of the data from the training set is randomly taken as the validation set. Our manual selection principle is to ensure that the correlation between the test data and the training data is minimal. By releasing this benchmark, we aim to establish a reliable reference for future research endeavors seeking fair comparisons in this field.

\begin{table}[ht]
\normalsize
    \caption{\label{tabD1} Train-test split for Shipsear. The ``ID'' in the table refers to the ID of the .wav file in the dataset.}
    \centering
	\scalebox{0.67}{
	\begin{tabular}{lll}
        \hline
	Category& ID in Training set & ID in Test set\\
	\hline
	Dredger&80,93,94,96& 95 \\
    Fish boat&73,74,76&75 \\
    Motorboat&21,26,33,39,45,51,52,70,77,79&27,50,72 \\
    Mussel boat&46,47,49,66&48 \\
    Natural noise&81,82,84,85,86,88,90,91&83,87,92\\
    Ocean liner&16,22,23,25,69&24,71\\
    Passenger ship&06,07,08,10,11,12,14,17,32,34,36,38,40,&9,13,35,42,55,62,65\\
    &41,43,54,59,60,61,63,64,67& \\
    RO-RO ship&18,19,58&20,78\\
    Sailboat&37,56,68&57\\
        \hline
	\end{tabular}}
\end{table}

\begin{table}[ht]
\normalsize
    \caption{\label{tabD2} Train-test split for DeepShip. The ``ID'' in the table refers to the ID of the .wav file in the dataset.}
    \centering
	\scalebox{0.69}{
	\begin{tabular}{lll}
        \hline
	Category& ID in Training set & ID in Test set\\
	\hline
    Cargo ship& Else & 01,02,04,05,18,30,32,35,40,48,56,62,\\
    & &63,67,68,72,74,79,83,91,92,93,95,97,\\
    & &100,104\\

    \hline
    Passenger ship& Else&02,04,05,07,11,15,17,19,20,23,30,31,\\
    &&37,45,46,52,53,60,61,62,64,67,68,70,\\
    &&75,76,77,84,86,91,101,106,113,117,122,\\
    &&125,129,130,134,135,142,144,152,157,\\
    &&159,161,167,168,177,179,187,188,189\\

    \hline
    Tanker& Else&02,03,04,07,08,13,14,15,19,22,25,28,\\
    &&35,37,46,58,62,71,73,79,82,84,88,89,\\
    &&92,99,106,115,118,124,126,127,131,134,\\
    &&141,144,147,151,153,156,158,167,171,\\
    &&178,179,185,186,190,192,193,201,205,\\
    &&213,217,228,233\\
    \hline
    Tugboat& Else& 07,08,18,20,24,25,27,29,32,33,37,39,\\
    &&40,44,45,56,59,70\\
        \hline
	\end{tabular}}
\end{table}

\subsection{Parameter Setup}
In this work, The frame length is set to 50ms and the frame shift defaults to half the frame length. We conduct experiments to investigate the effect of frame length on recognition results, as detailed in Section 5.1. Additionally, as a default, the number of Mel or Bark filter banks is set to 300.

During training, we employ the AdamW~\citep{loshchilov2017decoupled} optimizer with weight decay. The maximum learning rate is set to 5$\times 10^{-4}$, and the weight decay is set to $10^{-5}$ for all experiments. The models are trained for 200 epochs on A40 GPUs.

\section{Results and Analyses}
For the multi-class recognition task addressed in this work, we uniformly adopt the accuracy rate as the evaluation metric, which is determined by dividing the number of correctly predicted samples by the total number of samples. Besides, given the limited number of audio files in the test set, multiple groups of experiments yield the same file-level accuracy. Consequently, we present results at the segment level (30 seconds) rather than the file level. Moreover, to mitigate randomness, all reported results represent the average of experimental outcomes obtained using two distinct random seeds (42 and 123).

In this section, we initially conduct preliminary experiments on the frame length, effective frequency bands, and the structure of the front-end backbone network. After that, we perform main experiments to validate the effectiveness of CMoE based on four acoustic features and compare the experimental results with various advanced methods. This part also encompasses ablation experiments concerning the optional residual module and balancing regularization. We then provide a comprehensive visualization analysis of the expert assignment to further demonstrate the effective capture of useful information by the experts. Lastly, we carry out relevant experiments on the number of expert layers and the selection of the normalization function.

\subsection{Preliminary Experiments}

\begin{figure*}
    \centering
    \includegraphics[width=0.85\linewidth]{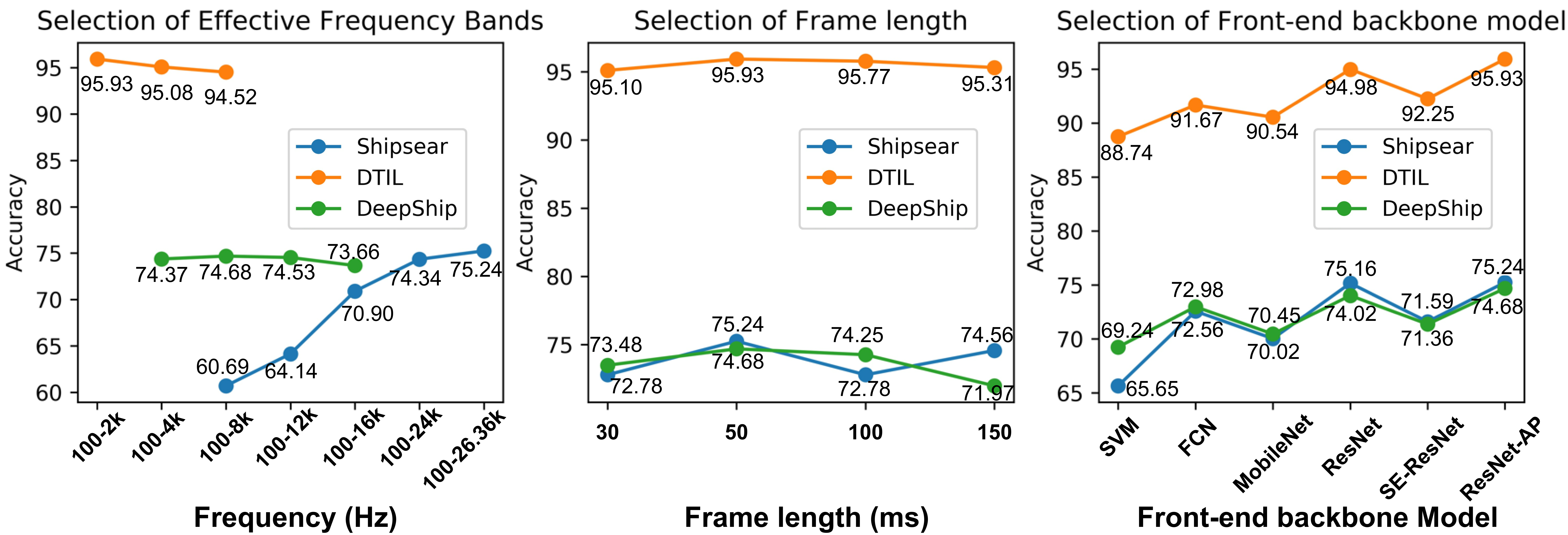}
    \caption{Preliminary experiments on the selection of effective frequency bands, frame lengths, and front-end backbone models. The frame shift is set to half the frame length by default.}
    \label{fig_pre}
    \vspace{-2px}
\end{figure*}



Before validating CMoE, we conduct preliminary experiments on the effective frequency bands, frame length, and structure of the front-end backbone network. The detailed results are presented in Figure~\ref{fig_pre}. Regarding the selection of effective frequency bands, relevant experiments consistently employ the STFT spectrogram as the input feature, utilize ResNet with attention pooling (ResNet-AP) as the model, and set the frame length to 50ms. A lower-cut-off frequency of 100Hz is set to filter out frequency bands with low signal-to-noise ratios. Experimental results indicate that Shipsear, DTIL, and DeepShip are suitable for utilizing 100-26.36kHz (Nyquist frequency), 100-2 kHz, and 100-8 kHz as effective frequency bands, respectively. For the selection of frame length, we also uniformly employ the STFT spectrogram as the input feature and ResNet-AP as the model. Optimal performance is consistently achieved with a frame length of 50ms across the three datasets. Therefore, a default frame length of 50ms is established.

Next, a comparison is made among six different backbone models - Support Vector Machine (SVM), Fully Convolutional Network (FCN), MobileNet-v3~\citep{howard2019searching}, ResNet~\citep{he2016deep}, SE-ResNet~\citep{hu2018squeeze}, and ResNet with attention pooling (ResNet-AP)~\citep{wang2018non} - for the selection of the backbone model. Among them, ResNet-AP demonstrates the highest recognition accuracy across the three datasets. Consequently, ResNet-AP is uniformly adopted as the default front-end backbone model in subsequent experiments.

\subsection{Main Results of CMoE}

\begin{table}[ht]
\normalsize
    \caption{\label{tab3} Main results of CMoE on four acoustic features. The ``Baseline model'' refers to the ResNet-AP selected in the preliminary experiments. Ablation experiments on the balancing regularization and the residual module are also included.}
    \centering
        \scalebox{0.65}{
	\begin{tabular}{llcccc}
        \hline
	Dataset & Model& STFT & Mel &Bark & CQT \\
        \hline
        Shipsear &AGNet~\citep{xie2022adaptive}& 
        85.48& - &- &- \\
          &Smooth-ResNet~\citep{xu2023underwater}& 
        81.90& 82.76 &- &75.86 \\
          &Baseline model& 
        75.24&77.14 &72.86 &73.33 \\
        &CMoE& 
        84.91&83.59&81.33&80.48 \\
          &CMoE+balance& 
        \textbf{86.21}&85.35 &84.48 &82.76 \\
          &RCMoE+balance& 
        85.34&84.48 & 83.62&82.76 \\
        \hline
        DTIL&AGNet& 
        95.76&- &- &- \\
          &TDNN\& WPCS~\citep{ren2019feature}&
        95.31&- &- &- \\
         &Baseline model& 
        95.93& 95.48&96.30 &96.48 \\
        &CMoE& 
        96.61&95.48&96.05&97.04 \\
          &CMoE+balance& 
        97.89& 97.46&96.89 &97.88 \\
          &RCMoE+balance& 
        \textbf{98.17}&97.60 &97.18 &97.89 \\

        \hline
        DeepShip&AGNet& 
        77.09&- &- &- \\
          &Smooth-ResNet~\citep{xu2023underwater}& 
        76.38& 77.05 &- &78.25 \\
         &SCAE~\citep{irfan2021deepship}& 
        -& 70.18&- &77.53 \\
         &Baseline model& 
        74.68&74.85 &75.15 &77.82 \\
          &CMoE& 
        75.65&76.09 &76.95 & 77.09\\
         &CMoE+balance& 
        76.33&76.72 &77.27 &\textbf{79.62} \\
          &RCMoE+balance& 
        76.80&76.60 &77.50 &78.76 \\
        \hline
    \end{tabular}}
\end{table}

Then, we experimentally validate the performance of our CMoE approach. To demonstrate its superiority, we compare our results with several existing advanced methods (AGNet~\citep{xie2022adaptive}, smoothness-inducing regularization~\citep{xu2023underwater}, TDNN and WPCS~\citep{ren2019feature}, SCAE~\citep{irfan2021deepship}) on three datasets\footnote{Since no official train-test split is released for these datasets, a fair comparison to substantiate the claim of being ``state-of-the-art'' is lacking. Therefore, instead of using the term ``state-of-the-art'', we refer to the selected methods as ``advanced methods''. We select these four methods for benchmarks as they have demonstrated competitiveness and superiority in their respective works.}. The main results of our experiments are presented in Table~\ref{tab3}. It is observed that vanilla CMoE can generally improve the recognition accuracy. Nevertheless, there are instances where the performance of CMoE may degrade due to the load imbalance issue, which is discussed in subsection 3.4. As an illustration, the CQT-based CMoE model experiences a 0.73\% decrease in accuracy compared to the baseline model on DeepShip. This decrease can be attributed to the under-training of certain experts caused by the load imbalance. Notably, this problem has the risk of worsening as the number of experts increases.

With the incorporation of balancing regularization, CMoE begins to demonstrate its potential. Particularly on Shipsear, where data originates from diverse regions and exhibits varied target motion statuses, the addition of the expert layers can help model learn effectively from diverse data and significantly increase the recognition accuracy from 75.24\% to 86.21\%. Furthermore, it is observed that CMoE proves effective across recognition systems based on all four features, confirming its generalizability.

Furthermore, we also conduct a set of ablation experiments on the residual module (see CMoE vs. RCMoE in Table~\ref{tab3}). It reveals that the residual module can not consistently yield improvements. RCMoE introduces a non-gated residual layer, which mitigates performance losses caused by inappropriate routing or expert overfitting. However, while addressing potential drawbacks, it also diminishes the sparsity of the network and reduces the disassembling effect of multiple expert layers on diverse data. Consequently, CMoE and RCMoE exhibit their respective strengths and weaknesses in different situations.

\begin{figure*}
    \centering
    \includegraphics[width=0.65\linewidth]{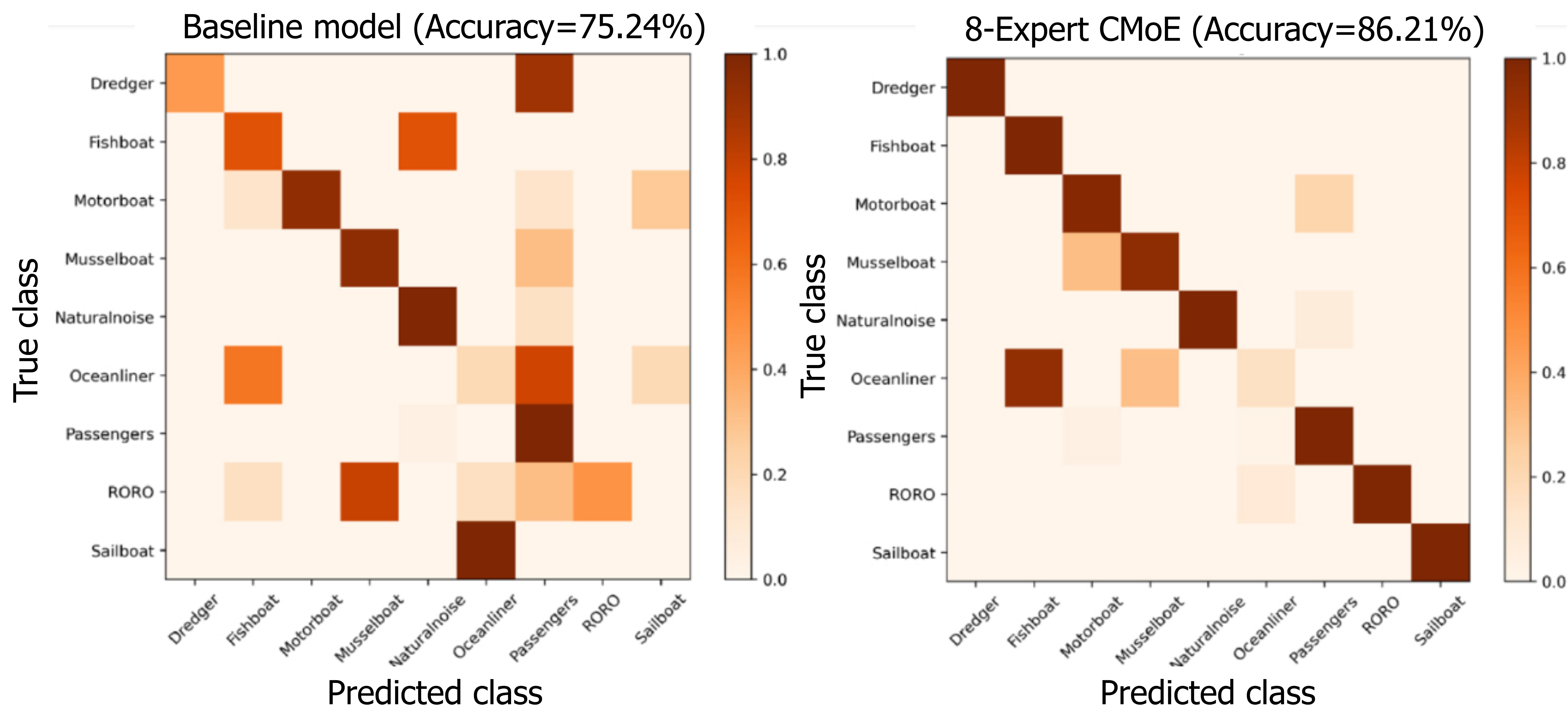}
    \caption{The confusion matrix heat maps of the baseline model and CMoE on Shipsear. Both models take the STFT spectrogram as the input feature.}
    \label{fig4}
    \vspace{-2px}
\end{figure*}

\subsection{Visualization Analyses}

We proceed to use the confusion matrix heatmaps to visually illustrate the model accuracy for each category. The confusion matrix compares the predicted labels (x-axis) with the actual labels (y-axis), and the heat map represents this matrix graphically, with colors indicating the values. In Figure~\ref{fig4}, it is evident that the baseline model's recognition performance is unsatisfactory, with some categories (dredger, ocean liner, ro-ro ship, sailboat) achieving accuracy rates below 50\%. For our proposed CMoE, the model achieves nearly 100\% accuracy for several categories that were previously challenging to recognize, such as dredgers, ro-ro ships, and sailboats, indicating a promising improvement.

\begin{figure*}
    \centering
    \includegraphics[width=0.65\linewidth]{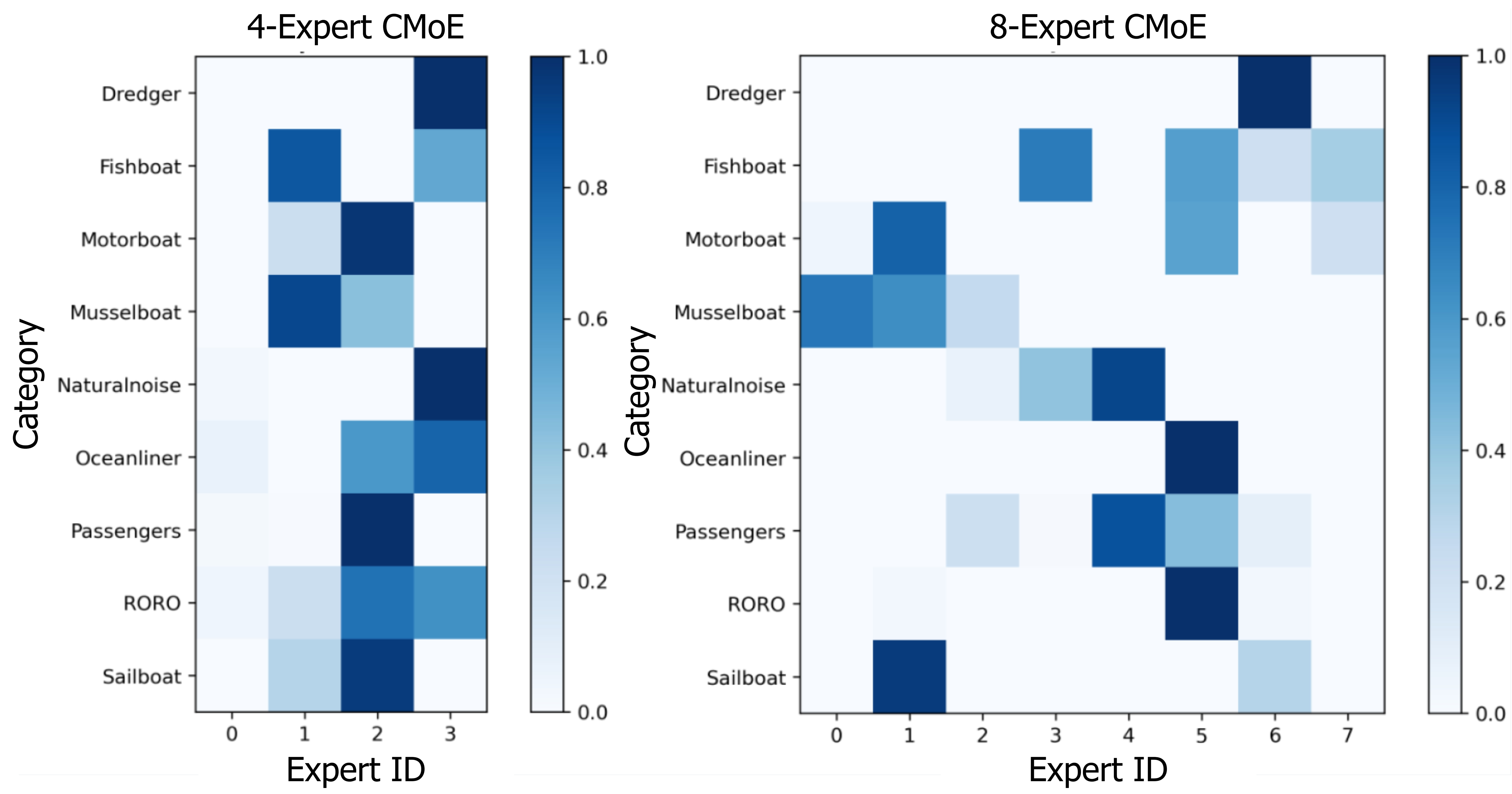}
    \caption{The overall assignment status of 4-expert CMoE and 8-expert CMoE on Shipsear. Both models take the STFT spectrogram as the input feature.}
    \label{fig5}
    \vspace{-2px}
\end{figure*}

Then, we perform a visualization analysis of the expert assignments in Shipsear. Heat maps are used to display the results, presenting the expert assignment for CMoE models with 4 and 8 experts (see Figure~\ref{fig5}). The colors in the heat maps represent the proportion of samples assigned to each expert category. Our analysis reveals a possible correlation between expert assignment and target size. According to experiential knowledge, ocean liners and ro-ro ships are classified as large targets, while motorboats and sailboats are considered small targets. As depicted in Figure~\ref{fig5}, the 8-expert CMoE tends to assign both ocean liners and ro-ro ships to the No.5 expert, while motorboats and sailboats are predominantly assigned to the No.1 expert. Similarly, the 4-expert CMoE exhibits similar assignments for small and large targets. This finding suggests that the routing module may capture inherent characteristics associated with target size.

\begin{figure*}
    \centering
    \includegraphics[width=0.7\linewidth]{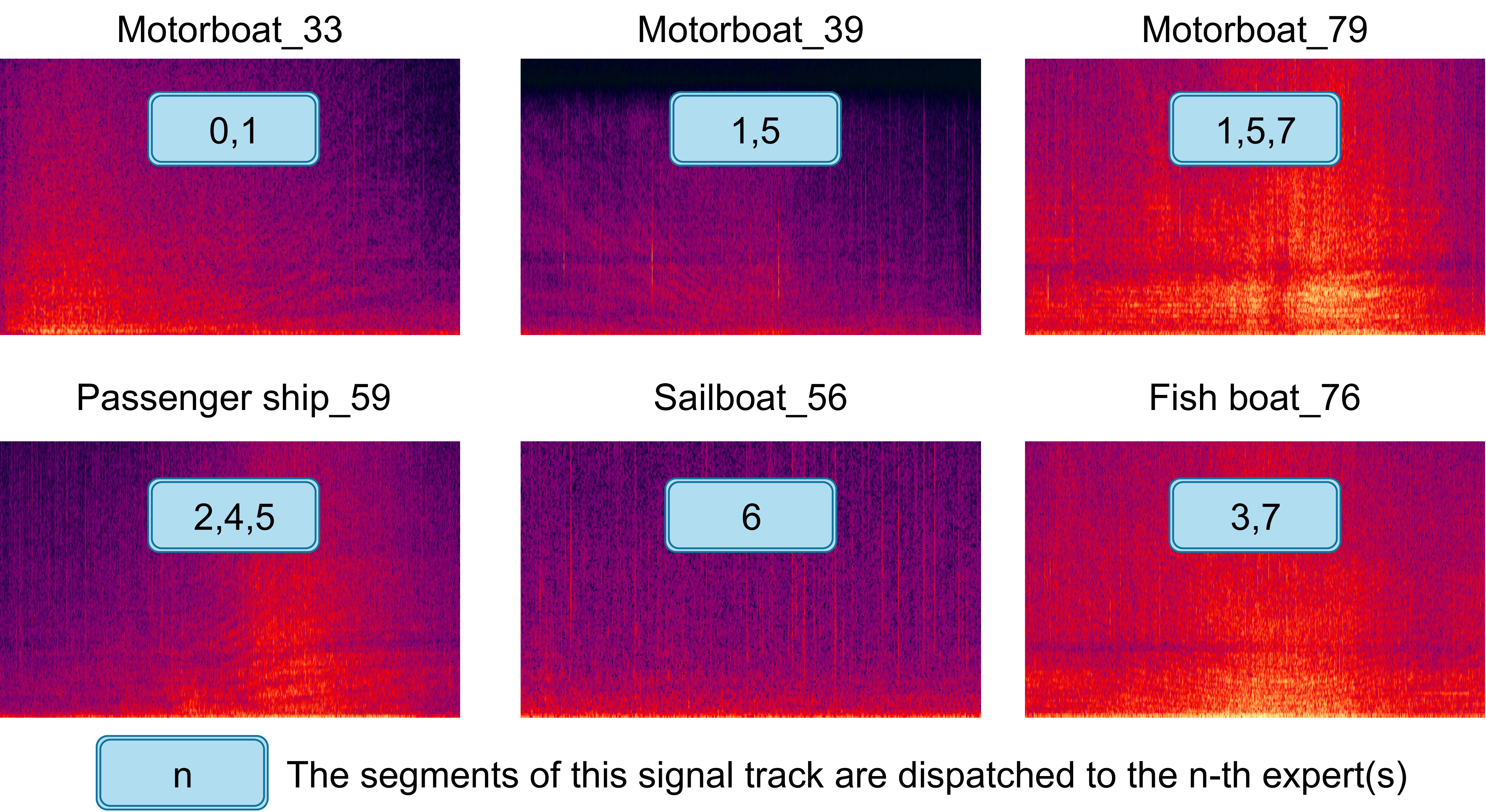}
    \caption{The expert assignment for the six samples. Since each sample can be cut into multiple 30s segments, these segments may be dispatched to multiple experts.}
    \label{fig7}
    \vspace{-2px}
\end{figure*}

Furthermore, Figure~\ref{fig7} presents the expert assignments for the six samples mentioned in Figure~\ref{fig1}. By comparing three motorboat samples with different motion states and types, we observe both overlaps and differences in their expert assignments. To some extent, the overlaps represent the target-related patterns, while the differences reflect the intra-class diversity. Additionally, we compare the expert assignment for samples from different categories with similar spectrograms (e.g., motorboat 33 vs passenger ship 59) and find that samples with similar spectrograms have different expert assignments. This suggests that our CMoE is successful in dealing with the inter-class similarity issue.


To conclude, the routing module responsible for expert assignment demonstrates the capability to capture target-related characteristics and intra-class diversity from high-level representations. Furthermore, we also find that the expert assignment of CMoE is little affected by inter-class similarity. The above visualization analysis provides certain interpretability for routing assignments.

\subsection{Number of Experts}
\begin{figure*}
    \centering
    \subfigure[Experiments on Shipsear.]{
        \begin{minipage}[b]{0.98\textwidth}
        \includegraphics[width=0.65\linewidth]{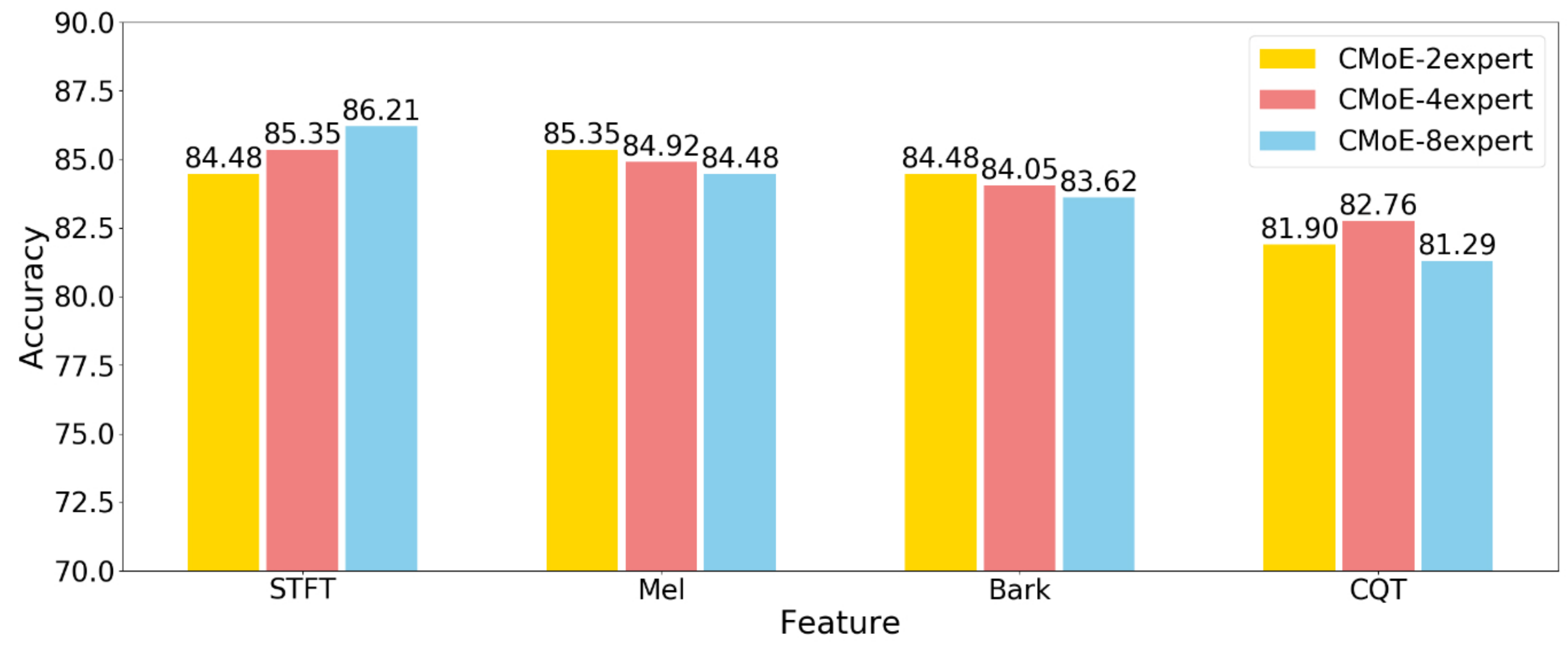}
        \centering
        \end{minipage}}

    \subfigure[Experiments on DTIL.]{
        \begin{minipage}[b]{0.98\textwidth}
        \includegraphics[width=0.65\linewidth]{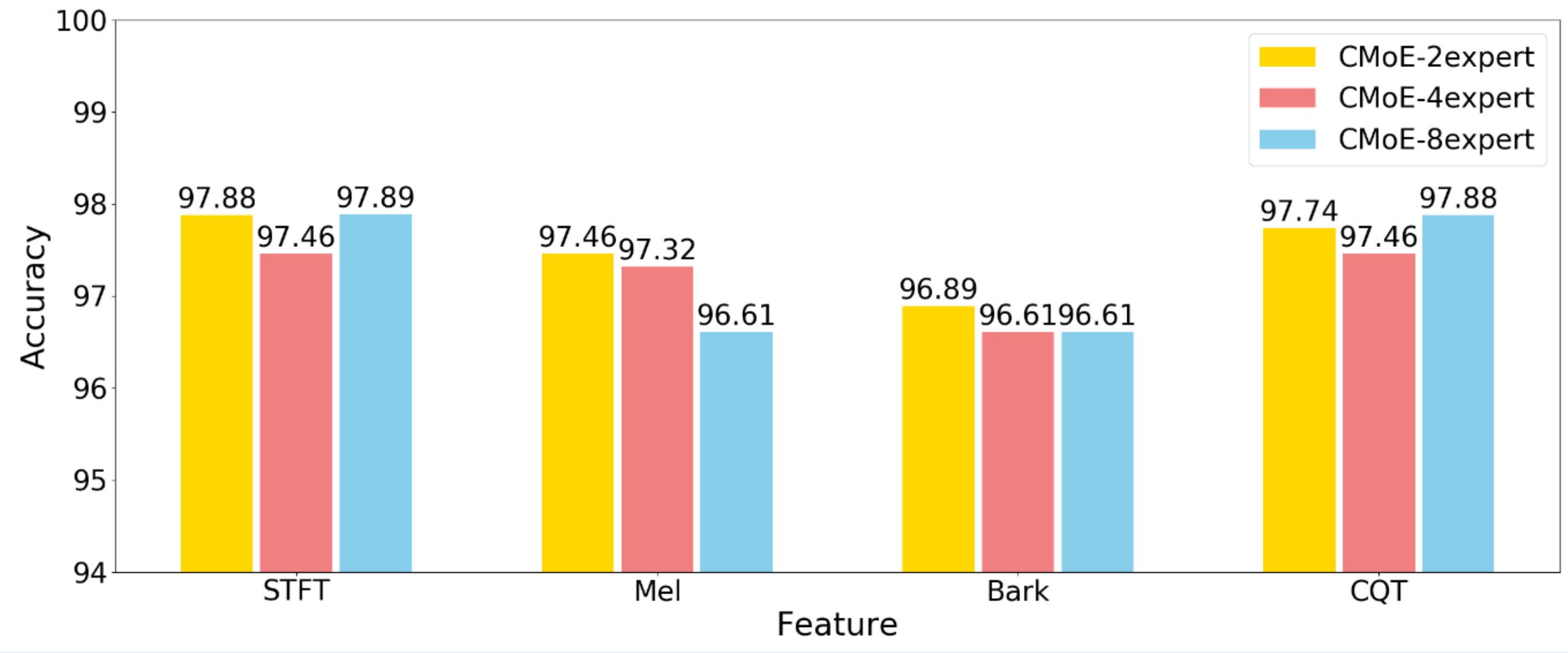}
        \centering
        \end{minipage}}
        
    \subfigure[Experiments on DeepShip.]{
        \begin{minipage}[b]{0.98\textwidth}
        \includegraphics[width=0.65\linewidth]{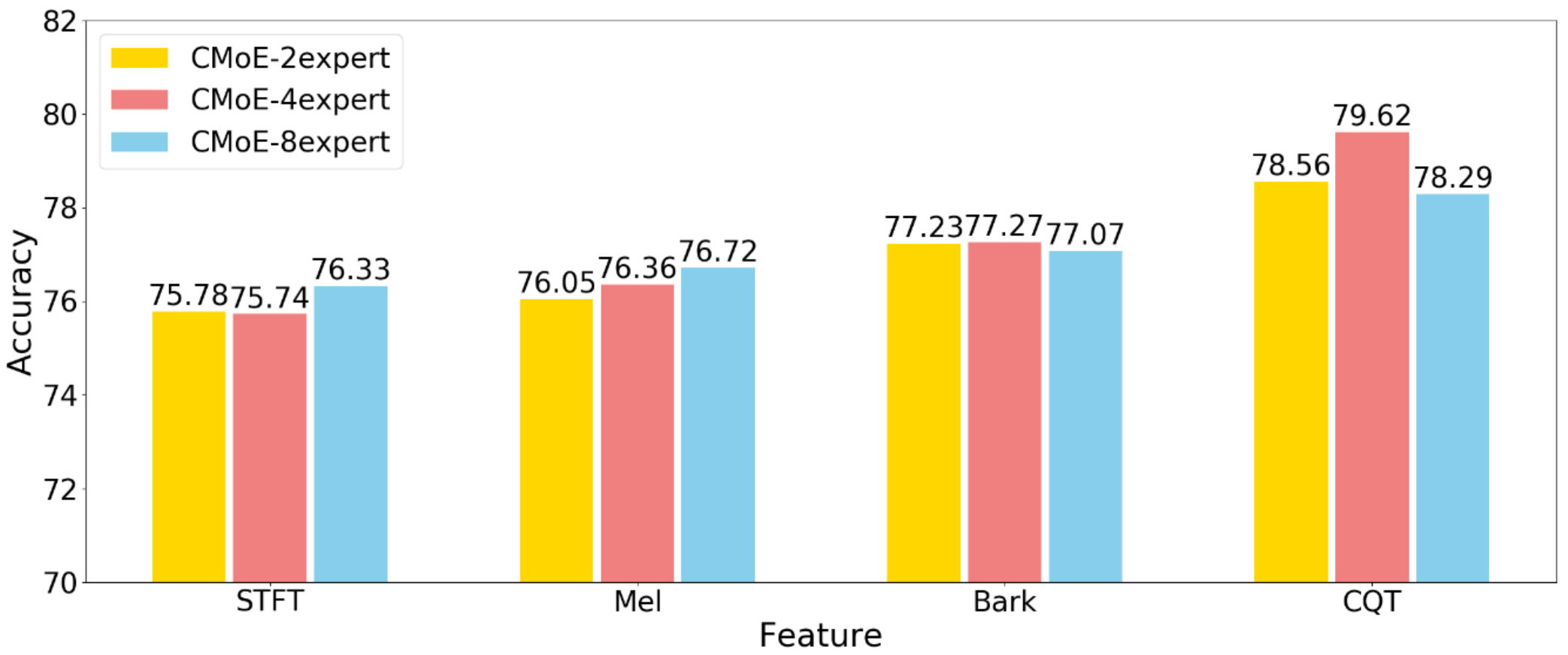}
        \centering
        \end{minipage}}
    \caption{Experiments about the number of experts with various features on three datasets.}
    \label{fig6}
    \vspace{-2px}
\end{figure*}

In this subsection, we delve into the examination of the influence of the number of experts in CMoE models. Our analysis reveals that while increasing the number of experts allows for a more fine-grained understanding of the data, it also diminishes the data dispatched to each expert, thereby increasing the risk of overfitting. Consequently, selecting the optimal number of experts becomes a complex trade-off. Figure~\ref{fig6} presents the results for CMoE models with 2, 4, and 8 experts across three datasets. The results illustrate that the impact of the number of experts is not simply proportional but varies depending on factors such as data scale, data diversity, input features, and other considerations. Only a portion of results show obvious correlations. On DeepShip, where the training data is relatively abundant compared to the other two datasets, CMoE with 2 experts exhibits the poorest performance. This is attributed to the limited capacity of CMoE to learn from diverse data. Besides, for redundant features, such as the STFT spectrogram on Shipsear with a dimension of 1200$\times$1318, increasing the number of expert layers can enhance performance. For other sets of experiments, the coupling of multiple influencing factors complicates the discovery of clear rules.


To summarize, selecting the optimal number of experts is a multifaceted trade-off task, and the impact of the expert layer on performance is contingent upon the characteristics of the data and features. Therefore, an extensive search for the optimal number of experts through traversal becomes necessary.

\subsection{Selection of the Normalization Function}

\begin{table}[ht]
\normalsize
    \caption{\label{tabB}  Experiments about the normalization function with various features on three datasets.}
    \centering
        \scalebox{0.7}{
	\begin{tabular}{lllcccc}
        \hline
	Dataset & Model&Norm func& STFT & Mel &Bark & CQT \\
        \hline
        Shipsear&CMoE&softmax& 
        84.48&83.19 &84.48 &82.76 \\
          &&sigmoid& 
        86.21&85.35 & 84.48&80.17 \\
          &RCMoE&softmax& 
        84.19& 82.76&83.19 &82.33 \\
          &&sigmoid& 
        85.34& 84.48 &83.62 & 82.76\\
        \hline
        DTIL&CMoE&softmax& 
        97.89&97.46 &96.89 &97.60 \\
          &&sigmoid& 
        97.88&96.89 & 96.61&97.88 \\
          &RCMoE&softmax& 
        98.17&97.18 &96.89 & 97.89\\
          &&sigmoid& 
        97.74& 97.60& 97.18& 97.32\\

        \hline
        DeepShip&CMoE&softmax& 
        76.33& 76.36& 76.52& 79.62\\
          &&sigmoid& 
        75.66&76.72 & 77.27&78.56 \\
          &RCMoE&softmax& 
        76.80&76.60 &77.50 &78.76 \\
          &&sigmoid& 
        76.23&76.46 &76.56 &77.99 \\
        \hline
    \end{tabular}}
\end{table}

Table~\ref{tabB} provides an overview of the impact of normalization functions on the recognition performance of CMoE and RCMoE. Note that the probability distribution to be normalized is $p$, and the formulas for the two normalization functions are as follows:

\begin{equation}
\begin{aligned}
    &Softmax(p_i)
    =\frac{e^{p_i}}{\sum_{j=1}^n e^{p_j}} (dim=-1),\\
    &Sigmoid(p)
    =\frac{1}{1+e^{-p}},
\end{aligned}
\end{equation}

where $p_i$ represents the probability distribution corresponding to the $i$-th class, while $n$ represents the total number of classes. The softmax function effectively converts the multi-category output values into a probability distribution within the range of (0, 1), where the sum of all probability values is equal to 1. On the other hand, the sigmoid function also maps the output values to the interval of (0, 1), with a steeper slope near the value of 0.5.

Upon analyzing the experimental results presented in Table~\ref{tabB}, it becomes apparent that the performance of the two normalization functions exhibits distinct advantages and disadvantages across different features and datasets. Although both functions only serve to normalize, they possess distinct gradient properties that can influence the weight updates of the network during backpropagation. Consequently, it is challenging to determine a universally superior normalization function. Both softmax and sigmoid normalization approaches are employed in this study, and the approach that yields superior results is selected.

\section{Conclusion}
This work unveils the uniqueness of underwater acoustic signals, characterized by high intra-class diversity and inter-class similarity. Building upon this foundation, we propose an innovative application of the mixture of experts to underwater acoustic recognition, called CMoE. This technique captures latent characteristics from high-level representations and adaptively learns diverse data with multiple independent parameter spaces. To optimize our model, we further incorporate balancing regularization and a residual module. Through comprehensive experiments, we demonstrate the superiority of our proposed method and underscore the necessity of balancing regularization. Furthermore, visualization analysis validates the effectiveness of our approach and enhances interpretability of the model.

Despite promising results, our CMoE still leaves certain limitations. First, we show that the assignment of experts is related to the intrinsic properties of targets (e.g., target size) to a certain extent, but the phenomenon lacks sufficient theoretical support. Moreover, the design of the structure of expert layers and routing layer is relatively simple. We believe that the current CMoE with simple linear layers as the experts or the routing layer does not fully exploit the potential of the MoE structure. In our future work, we aim to investigate the potential of utilizing physically-based target characteristics, such as the number of propeller blades, as the foundation for routing instead of relying solely on the automatically learned routing layer. This approach offers improved interpretability and holds promise for our research endeavors.

\section*{Acknowledgements}
This research is supported by the IOA Frontier Exploration Project (No. ZYTS202001) and Youth Innovation Promotion Association CAS.

\newpage

\printcredits

\bibliographystyle{cas-model2-names}

\bibliography{cas-refs-cmoe}





\end{document}